# What Smartphones, Ethnomethodology, and Bystander Inaccessibility Can Teach Us About Better Design?

Eerik Mantere[1,2]

[1] Tampere University, Kalevantie 4, 33100 Tampere, Finland
[2] Université de Bordeaux, 3 ter Place de la Victoire, 33076 Bordeaux, France
emantere@u-bordeaux.fr

**Abstract.** Smartphones, the ubiquitous mobile screens now normal parts of everyday social situations, have created a kind of ongoing natural experiment for social scientists. According to Garfinkel's ethnomethodology social action gets its meaning not only from its content but also through its context. Mobility, small screen size, and the habitual way of using smartphones ensure that, while offering the biggest variety of activities for the user, in comparison to other everyday items, smartphones offer the least cues to bystanders on what the user is actually doing and how long it might take. This 'bystander inaccessibility' handicaps shared understanding of the social context that the user and collocated others find themselves in. Added considerations and interactive effort in managing the situation is therefore required. Future design needs to relate to this basic building block of collocated interaction to not be met with discontent.

## 1 Introduction and Background

In United States 81% owns a smartphone [1] and they are routinely used in the presence of others. How this impacts relationships with collocated others regularly hits the headlines [2][3]. Previous research suggests various negative effects. Smartphone use can be distracting and undermine the benefits of social interactions [4], which have previously found to be so crucial to psychological well-being [5]. Although often aiming for connection with distant others, interactions online do not provide the same sense of social support as collocated interactions [6]. Being distracted in collocated interactions due to smartphone use therefore seems like an ill-chosen trade-off.

An Australian dictionary jumped on the idea by coming up with a new word for the phenomenon. "Phubbing" is defined at their marketing campaign's website site as *"the act of snubbing someone in a social setting by looking at your phone instead of paying attention"* [7]. Researchers got on board with the term and phubbing has since been found to reduce communication quality and relationship satisfaction by reducing the feelings of belongingness and positive affect [8], make both phubbers and the phubbed to be more likely to see phubbing as an inevitable social norm [9], and be thought of as 'bad' by young people, even if they are doing it themselves [10]. "Partner phubbing" has further been found to reduce relationship satisfaction by creating conflicts over cell phone use [11] and cause depression in China for couples married more than seven years [12]. A validated scale to measure phubbing has also been developed [13] and the capacity to predict phubbing risk has been pursued by forming a model constituting of communication disturbances and phone obsession [14]. One should not then be surprised then that an article in the New York Times portrayed phubbing as if the term was developed by psychologists [15].

Not wanting to discredit the previous work, three points should be noted of their similar methodologies and the gap they fail to fill. First, though they study the social situation, they do not directly describe it, but rather produce second level constructs of it [16]. Research participants have produced numeric or verbal accounts of imagined or previously lived situations. These are then used to make a scientific accounts—now two levels distanced from the phenomenon they aim to depict. Second, when directly observing social situations, they rely on *a priori* chosen qualities of interaction. Researchers observing social behavior then code it in regard to these qualities in order to use them as indicators in seeking relevance between them and general social categories like age or gender [25]. Third, none of them spring from a theoretical understanding of social action. Harold Garfinkel pointed out the problems of theories that rely on internalization of society's norms and found ethnomethodology (EM) to study how people themselves in everyday situations construct meaning and make and interpret social typifications as relevant. EM has quickly gained more and more ground as *the* theory of social action and has given birth to conversation analysis (CA), now considered the principal way to study verbal and non-verbal interaction alike [16] [17] [18] [19] [20].

Though EM/CA literature covers a wide range of interactive contexts, research on spontaneous individual smartphone use in social situation is practically non-existent. One of the most closely related EM/CA studies looked at how smartphone use while driving is interleaved with traffic light stops. Users were looking for moments when the affordances of the phone's interface co-constructed transition relevant places with the activities of the user. In these moments a possible shift in orientation between smartphone use and other activities is sequentially made most available. The regularity in which the interface makes these moments possible was considered a central theme in organizing multiactivity with smartphone use and other concurrent activities [21]. Another study of using public transport found gaze shifts away from the phone to be organized in relation to the sequential progression of the activity with the device. Beginning stages of phone use were suggested to be especially sequentially engaging but the methodology used and the level of granularity of the analyzes lacked the possibility to describe the interactive practices of in their sequential contexts [22].

A study focusing on the use of a map applications found people sequentially organizing their phone use with actions like unilateral stopping, turning, and restarting, while walking together in public places. Again, phone use was found to have its own sequential progression which, then was interleaved with that of the concurrent social activities of the physical environment [23]. The most relevant EM/CA work on smartphone use and collocated interaction addresses phone use in pubs [24]. It does introduce and explore the topic but does not exhaust neither a single episode of interaction, nor describe any putative practice taking place in various interactions, to a satisfactory degree from the point of view of CA. Similarly, it does not make real use of the theoretical offer of EM. I encourage looking at smartphone use in social situation with a viewpoint rooted in EM, and adding in CA analysis, in order to understand how phone use may be constructed as unacceptable, and to find inspiration for more socially acceptable design.

## 2 Social Theory and Indexicality

Goffman [25] defined the social situation as an *„environment of mutual monitoring possibilities, anywhere within which an individual will find himself accessible to the naked senses of all others who are ¨present,¨ and similarly find them accessible to him.“* All speaking and gesturing in face-to-face interaction takes place in the social situation and he emphasized the importance of the physical setting in any analysis of them. Even more than Goffman, Harold Garfinkel saw the context of interaction to be central in what the interaction itself means [16] [18]. Let us consider the following example:

```
I'm sorry
```

The phrase seems to clearly convey an apology. We might imagine that the person uttering the words feels regretful and elucidate how each of the words, I – am – sorry, convey something that together constitute an apology. We might reflect on how it differs from the more casual "sorry 'bout that", and we might even say that this apologizing seems humble and empathetic. But what if we added a context:

```
I got my diploma from the University of Honolulu

I'm sorry
```

Now the phrase "I'm sorry" doesn't seem so kind. This example shows how the same practice of "apologizing" can be used to do different things—one of them teasing. As the immediate social context changes, the meaning of the action changes too. Before Garfinkel, 'indexicality' was considered as a character only of words like "this", "here" or "now"—words that point, or index, a context in acquiring their meaning. Garfinkel planned a series of breaching experiments to claim that actually all human action is understood as indexing the context it takes place in. If people encounter behavior that is not designed in relation to the commonly shared situation, they feel awkward and severely challenged in knowing how to proceed. Whatever is done, through words or otherwise, always gets interpreted through what is seen as the *shared understanding* of the situation that the action takes place in [16] [18].

Garfinkel further proposed that this understanding was not only his, but people conducting their everyday lives actually orient to each other as *accountable* in entering social situations with the assumption that it is common knowledge [18]. This knowledge is not rooted in detached reflection of the deep nature of social action. He does not suggest that all members of society passed sleepless nights in understanding the core concepts of ethnomethodology. Rather, in interacting with one another, a general thesis of interchangeability of perspectives is at work. To put it simply, people assume that what they see as relevant in a situation is seen relevant by others in the same situation. This is crucial for being able to trust to the shared understanding of the social situation as “good enough” for interaction to be meaningful. If we could not trust that we and another person have at least “good enough” match in understanding what is going on, we could not trust that anything we say or do in the situation would be understood as we would like it to be understood.

## 3 Bystander Inaccessibility

The participants of a social situation who start to use a smartphone, to a large extent stop giving hints of the goals of their actions to collocated others. Others can less often than is the case with other devices, infer from the posture and movements of the user, or from the shape and state of the smartphone, what the user is currently doing. The lack of visual and auditive cues to the bystander, the mobility of the device, bigger amount of variation in the types of actions possibilitie, than is the case with any other device, and the varied temporal organizations of the different smartphone activities are responsible of keeping some crucial aspects about the smartphone user’s activity hidden to the person in their immediate vicinity:

I. Phase of action (e.g. preparatory phase, execution phase, or being already close to terminating the action)

II. Category of action (e.g. entertainment, work, information seeking, or communication)

Not knowing what the activity of the smartphone user is, the other participants in the social situation are also in the dark about the “good enough” knowledge about nature of the situation as a whole. I call this bystander inaccessibility (BI). Imagine you want to ask something mundane of your partner, let’s say, if she has gotten the mail. The mailbox is just outside, and you could easily check it yourself, but you would prefer not getting out of the house in vain. You see your partner sitting on the sofa, absorbed in their phone. Now if you would know that they are responding to an important work email, you might leave them alone and check the mail yourself. But if they were just scrolling a friend’s Facebook feed, you might feel at ease to interrupt them. Being in the dark about the activity they are engaged in, you are also unable to know what your planned communicative action, “have you checked the mail?”, would signify to them.

It works the other way around as well. This is exemplified in the following data excerpt. Clo and Liz are eating out and exchanging funny stories together with a friend.

Excerpt 1.

[overlapping speech]

>faster speech<
(0.9) silence in tenth of seconds
(.) noticeable silence of maximum 0.3 seconds
.mt smacking of the mouth
@transformed speech, e.g. when quoting someone@
°spoken silently°
the- the production of the word is halted suddenly
((comments))

((Clo is using her phone while talking))
64 Clo: [>Nii nimeonomaa<] (.) ja sit vielä se ku tota noin ni toi
*[>Yeah exactly<] (.) and then also that when you know that*

65 (0.9) .mt ((Clo stops typing and puts left hand to her face))

66 (0.2) ((Clo continues to gaze at the phone))

67 Clo: öö iskä >oli sillee< [@↑nii joo mä muistan kun Niina
*umm dad >was like< [@oh yeah I remember when Niina*
68 Liz: [°mä katon ton-°
*[°I'll check th-° ((picks up her phone))*

Clo is starting to tell a story that continues the theme of previous stories that night. While doing this she pauses (line 65, 66) and utilizes filler words (lines 64–67) before actually getting the story going (line 67). Before her turn she was using her phone. While beginning the telling at line 67, she is still looking at it. Liz is listening, gazing at Clo, and sees all this taking place. While Clo is struggling while visibly distributing her attention between two activities—telling a story and using her phone—she is also putting Liz in a difficult position. Clo has already prefaced her story and gained a silent "permission" from the group to occupy a speaker position for a longer duration than normal, i.e. until the story is finished. Therefore other participants are normatively restricted to the position of recipient. When regardless of this, Clo still does not put her full attention to the activity of telling the story, and is faltering in beginning the story, the next activities, being indexical, connect in their meaning also to this event.

When Liz begins to use her phone at line 68, BI makes Clo unable to automatically see the type and the goal of Liz's phone use. In this context it therefore risks being interpreted as motivated by dissatisfaction with the haphazard way Clo begun her responsibilities as a storyteller. Considering Goffman's [26] face-work and the normative ways we protect the faces of ourselves, as well as other people from straightforward criticism, it is understandable that Liz chooses to counter this potentially face-threatening interpretation. She provides an account: "I'll check the-" at line 68. Interestingly, she does not actually specify the activity she will commence with the phone, but in providing the account, she nevertheless hints that there is something to be "checked" and the reason for her staring to use the phone could be in this "checking", rather than in the faltering conversational performance of Clo. To conclude, as BI hides Liz's activity from Clo, Liz has to produce an account to circumvent this lack. Providing this account in a sequentially appropriate manner encumbrances a very limited resource in the context of being a recipient to verbal storytelling: audible speech.

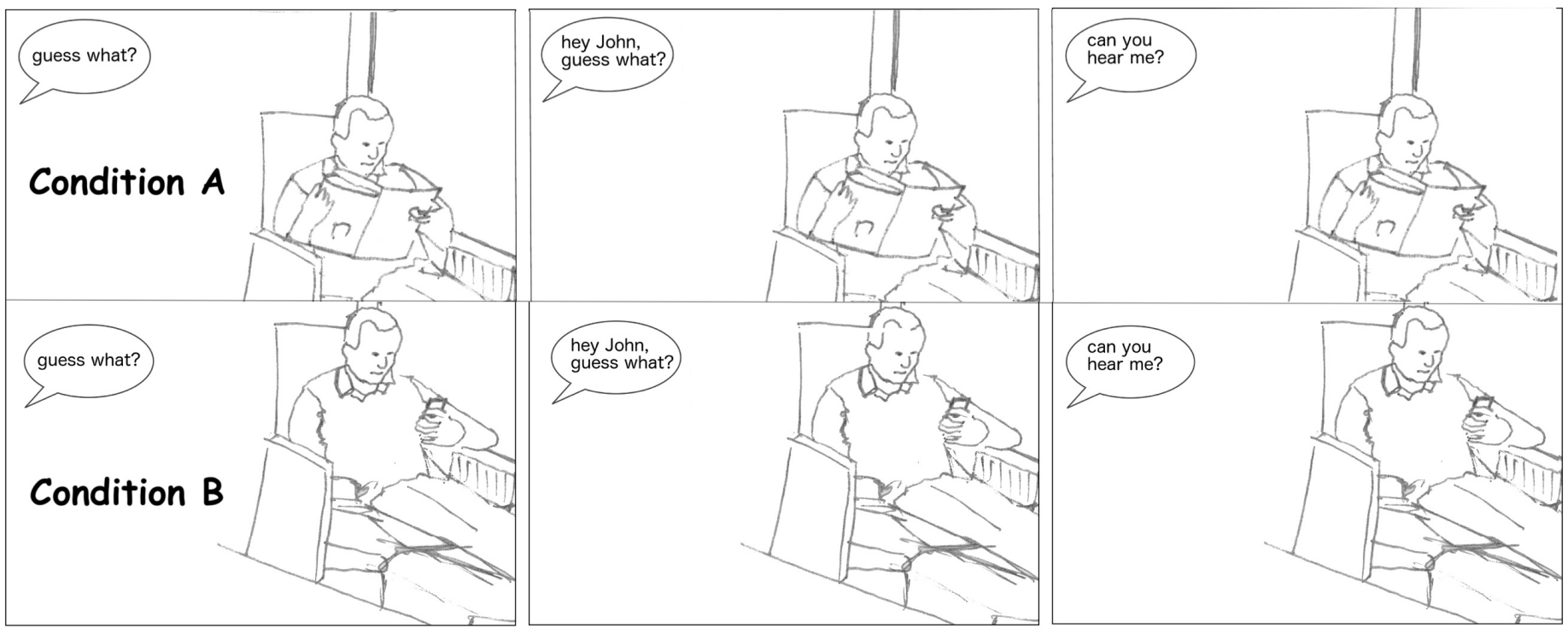

**Fig. 1.** Respondents identified with the person speaking and rated A and B in random order.

BI -instigating technology (BI-tech) also makes it harder for collocated others to interpret responses, or lack of them, by a BI-tech user. Our study using role playing method and comic strips found most respondents more irritated when trying to unsuccessfully get the attention of a phone-using person, while no respondents evaluated the newspaper -condition as more irritating ($p<0.001$). Furthermore, the written responses often included descriptions on being bothered by not knowing what the phone user was actually doing [27].

## 4 Conclusion

Designing socially acceptable technology should be informed by ethnomethodological study on the device's effect on social situation. What people do or do not accept is the way technology enters into the situation as part of the network of social activities. When engaged in technology use, a crucial aspect of it is that the activity is part of constituting the shared social reality that then gives meaning also to all the other activities of everyone else present in the situation. All their decisions to act or not to act are impacted by their understanding of what the technology use is about and whether they can trust that other participants see the situation similarly. There should be more work on design instigating affordances for collocated others to see, hear, or feel the nature of the technology use taking place in a social setting [28]. Crucially, I call for interdisciplinary work that benefits from EM/CA methodology to develop and test new prototypes. BI-tech handicaps participants in social encounters. While people find ways to circumvent it, the plethora of research reporting dislike of smartphone use in social situation suggests they would prefer to avoid these challenges. Interactional work and designing non face-threatening actions takes effort, and people do not like to be forced to make effort.

## References

1. Taylor, K., & Silver, L.: Smartphone Ownership Is Growing Rapidly Around the World, but Not Always Equally. Pew Research Center, (2019).
2. Ducharme, J.: "Phubbing" Is Hurting Your Relationships. Here's What It Is. TIME, (2018).
3. Molina, B.: Do smartphones keep us in or out of touch?: Devices often isolate, distract and disrupt acting with others. USA TODAY, (2017).
4. Dwyer, R. J., Kushlev, K., & Dunn, E. W.: Smartphone use undermines enjoyment of face-to-face social interactions. Journal of Experimental Social Psychology 78, 233–239 (2018).
5. Feeney, B. C., & Collins, N. L.: A New Look at Social Support: A Theoretical Perspective on Thriving Through Relationships. Personality and Social Psychology Review 19(2), 113–147 (2014).
6. Kim, J.-H.: Smartphone-mediated communication vs. face-to-face interaction: Two routes to social support and problematic use of smartphone. Computers in Human Behavior 67, 282–291 (2017).
7. Stop Phubbing Website, http://stopphubbing.com, last accessed 2019/3/22.
8. Chotpitayasunondh, V., & Douglas, K. M.: The effects of "phubbing" on social interaction. Journal of Applied Social Psychology 48(6), 304–316 (2018b).
9. Chotpitayasunondh, V., & Douglas, K. M.: How "phubbing" becomes the norm: The antecedents and consequences of snubbing via smartphone. Computers in Human Behavior 63, 9–18 (2016).
10. Aagaard, J.: Digital akrasia: a qualitative study of phubbing. AI and Society, 1–8 (2019).
11. Roberts, J. A., & David, M. E.: My life has become a major distraction from my cell phone: Partner phubbing and relationship satisfaction among romantic partners. Computers in Human Behavior, 54(Journal Article), 134–141 (2016).
12. Wang, P., Wang, X., Wang, Y., Xie, X., & Lei, L.: Partner phubbing and depression among married Chinese adults: The roles of relationship satisfaction and relationship length. Personality and Individual Differences 110, 12–17 (2017).
13. Chotpitayasunondh, V., & Douglas, K. M.: Measuring phone snubbing behavior: Development and validation of the Generic Scale of Phubbing (GSP) and the Generic Scale of Being Phubbed (GSBP). Computers in Human Behavior 88, 5–17 (2018a).
14. Guazzini, A., Duradoni, M., Capelli, A., & Meringolo, P.: An explorative model to assess individuals' phubbing risk. Future Internet 11(1), 21–34 (2019).
15. Roose, K.: Do Not Disturb: How I Ditched My Phone and Unbroke My Brain. New York Times (2019).
16. Garfinkel, H.: Studies in ethnomethodology. Prentice-Hall, Englewood Cliffs, N.J. (1967).
17. Goodwin, C.: Conversational organization: interaction between speakers and hearers. Academic Press, New York (1981).
18. Heritage, J: Garfinkel and Ethnomethodology. Polity Press, Cambridge (1984).
19. Hutchby, I., & Wooffitt, R.: Conversation analysis: principles, practices and applications. Polity Press, Cambridge (1998).
20. Mondada, L: Multimodal resources for turn-taking: pointing and the emergence of possible next speakers. Discourse Studies 2(9), 194–225 (2007).
21. Licoppe, C., & Figeac, J.: Gaze Patterns and the Temporal Organization of Multiple Activities in Mobile Smartphone Uses. Human-Computer 33(5–6), 311–334 (2018).
22. Figeac, J., & Chaulet, J.: Video-ethnography of social media apps' connection cues in public settings. Mobile Media & Communication 6(3), 407–427 (2018).
23. Laurier, E., Brown, B., McGregor, M.: Mediated Pedestrian Mobility: Walking and the Map App. Mobilities 11(1), 117–134 (2016).

24. Porcheron, M., Fischer, J., & Sharples, S.: Using Mobile Phones in Pub Talk. Proceedings of the 19th ACM Conference on Computer-Supported Cooperative Work & Social Computing, 1649–1661. ACM (2016).
25. Goffman, E.: The Neglected Situation. American Anthropologist 66(6), DEC–136. (1964).
26. Goffman, E.: Interaction ritual: essays on face-to-face behavior. Aldine, Chicago (1967).
27. Raudaskoski, S., Mantere, E., & Valkonen, S.: Älypuhelin ja kasvokkaisen vuorovaikutuksen muuttuvat käytänteet. *Sosiologia* (accepted for publication) (2019).
28. Ens, B., Grossman, T., Anderson, F., Matejka, J., & Fitzmaurice, G.: Candid Interaction: Revealing Hidden Mobile and Wearable Computing Activities. Proceedings of the 28th Annual ACM Symposium on User Interface Software & Technology, 467–476. (2015).